\documentclass[preprint,superscriptaddress,showpacs]{revtex4-1}
\usepackage{epsfig}
\usepackage{graphics,graphicx}
\usepackage{amssymb}
\usepackage{mathrsfs}
\usepackage{amsmath}
\usepackage{verbatim}
\usepackage{hyperref} 
\usepackage{rotating}
\usepackage{float}
\usepackage{setspace}

\begin{document}

\title{Effects of confinement between attractive and repulsive walls
  on the thermodynamics of an anomalous fluid} 

\author{Fabio Leoni}
\affiliation{Secci\'o de Física Estad\'istica i
Interdisciplin\`aria - Departament de F\'isica de la Mat\`eria
Condensada, Facultat de F\'isica, Universitat de Barcelona, Mart\'i i
Franqu\`es 1, Barcelona 08028, Spain}
\author{Giancarlo Franzese}
\affiliation{Secci\'o de Física Estad\'istica i
Interdisciplin\`aria - Departament de F\'isica de la Mat\`eria
Condensada, Facultat de F\'isica, Universitat de Barcelona, Mart\'i i
Franqu\`es 1, Barcelona 08028, Spain}
\affiliation{Institut de Nanoci\`encia i Nanotecnolog\'ia, Universitat
  de Barcelona, Av. Joan XXIII S/N, Barcelona 08028, Spain}
 
\begin{abstract}

We study by molecular dynamics simulations the thermodynamics of  an
anomalous fluid confined in a slit pore with one wall structured and
attractive and another unstructured and repulsive.
We find that the phase diagram of the homogeneous part of
the confined fluid is shifted to higher temperatures, densities and 
pressures with respect to the bulk, but it can be rescaled on the bulk
case. 
We calculate a moderate increase of mobility of the homogeneous
confined fluid that we interpret as a consequence of the
layering due to confinement and the collective modes due to long-range
correlations.
We show that, as in bulk, the confined fluid has structural, diffusion
and density anomalies, that order in the water-like hierarchy, and a
liquid-liquid critical point (LLCP).
The overall anomalous region moves to higher temperatures, densities
and pressure and the LLCP  displaces to higher temperature compared to
bulk.  
Motivated by experiments, we calculate also the phase diagram not just
for the homogeneous part of the confined fluid but for the entire
fluid in the pore and show that it is shifted towards higher pressures
but preserves the thermodynamics, including the LLCP.
Because our model has water-like properties, we argue that in
experiments with supercooled water confined in slit pores with a width
of $>3$~nm if hydrophilic, and of $>1.5$~nm if hydrophobic, the
existence of the LLCP could be easier to test than in bulk, where it
is not directly accessible.
\end{abstract}

\maketitle

%
%

\section{Introduction}
\label{sec:introduction}

Many experiments and simulations have shown that fluids under
geometrical nano-confinement present a different structural, dynamical
and thermodynamic behavior with respect to the bulk case.
The study of such systems is relevant for theirs technological,
experimental, and theoretical implications \cite{Bellissent95,
schoen1998, Mashl2003, Mittal2008, cicero2008, De-Virgiliis2008,
Giovambattista09, castrillon2009, castrillon2009b, Mancinelli2009,
Rzysko2010, Han2010, delosSantos2011, schnell2011, schnell2012,
Nair2012, Paul2012, Ferguson2012, Stewart2012, Krott2013,sun2015,
klein1995, thompson1992, cui2001, jabbarzadeh2007, ramin2013,
lorenz2009}.  
Despite these many experimental and computational studies, the effects
of a specific nano-confinement on the properties of a fluid are still
under debate. 
It has been shown that walls with different characteristics, but in the
same confinement geometry, have drastically different effects on
crystal nucleation and fluid dynamics.
Because in some case experiments are difficult to interpret and
atomistic simulations require elaborated approaches to exclude
numerical artifacts, effective potentials have been employed to study
nano-confined fluids. 
They have the advantage to be much less computationally expensive than
atomistic simulations and, even though they can not give quantitative
interpretation of experiments, they can allow us to understand the
mechanism behind the effect of confinement on the fluid. 
 
Here we consider a system of particles interacting through the
Continuous Shouldered Well (CSW) effective potential
\cite{Fr07a,OFNB08,vilaseca:084507,leoni2014} confined in a slit pore.
The CSW is an isotropic pairwise core-softened potential with a
repulsive shoulder and an attractive well
\cite{Fr07a,OFNB08,vilaseca:084507}, similar to other
used for colloids \cite{Ryltsev:2015aa,  Denton:1998aa} and water-like liquids
\cite{Yan:2008ve,Agarwal:2010vn,Krott2015,sun2015}.
It is suitable for describing
globular proteins in solution \cite{C3SM50220A},
methanol hydroxyl
groups \cite{Hus:2014aa,Hus2014}, liquid metals \cite{Vilaseca2011},
water-like liquids \cite{OFNB08} and can properly describe the
hydrophobic effect of water as a solvent \cite{Hus2013}. 
In particular, the CSW model reproduces density, diffusion, and
structure anomalies following the water hierarchy and displays a
gas-liquid and a liquid-liquid phase transition (LLPT), both ending in
critical points \cite{OFNB08,vilaseca:084507}.

We adopt a slit pore confinement with a separation between the two
walls of the order of ten layers of the fluid \cite{leoni2014}. This
pore size  is particularly important in nanotribology and nanofluidics
of thin water films, for which there are contradictory results from
measurements of the dynamical and structural properties when
nanoconfined in hydrophilic and hydrophobic slit pores \cite{Khan2010}.

Our  slit pore is composed by an attractive wall with atomistic
structure and a smooth repulsive wall. This confinement allows us to
investigate at the same time the effect of both surface interactions
and structures.
An asymmetric (semihydrophilic) slit pore, similar but opposite to
ours, is adopted in experimental studies of water nano-confined
between a smooth (hydrophilic) mica wall and a structured
(hydrophobic) graphene sheet \cite{severin2012}.
A similar confinement has been also considered in previous theoretical
works, e.g., in relation with wetting phenomena
\cite{parry1990,albano2012,swift1991}. 
Smooth repulsive walls have been used as coarse-grained models for
hydrophobic surfaces \cite{Liu2004,Krott2015}.

Our motivation follows from our previous work \cite{leoni2014} in
which we find layering of the CSW particles, as expected. In
particular, we observe that the structured attractive wall has an
effect on the fluid that is stronger than the smooth repulsive wall. 
At low temperature $T$ we observe heterogeneous crystallization
starting at the attractive wall. The crystal propagates toward the
repulsive wall as $T$ decreases and density $\rho$ increases.
At low $T$ the large $\rho$ of the attractive
wall induces a high-density, high-energy structure in the first layer
(``templating'' effect). In turn, the first layer induces a ``molding''
effect on the second layer determining a structure with reduced energy
and density, closer to the average density of the system. This
low-density, low-energy structure propagates further through the
layers by templating effect and can involve the entire system at low $T$.
By increasing $T$ the crystal layers near the walls became 
amorphous and those in the center of the slit pore become liquid.
Furthermore, at low $T$ and high $\rho$  we observe
that the dynamics is largely heterogeneous ``among'' and ``within''
the layers, with the possible formation of superdiffusive
liquid veins within the crystal. 

We find that the homogeneous fluid at the center of the slit pore has
a phase diagram with the same features as the bulk: a gas-liquid
critical point (GLCP), a liquid-liquid critical point
(LLCP), and a water-like cascade of
anomalies, with the structural-anomaly region, the
diffusion-anomaly region and the density-anomaly region including one
the other in succession. However, we find that the
confinement strongly enhances the structure and accelerates the
dynamics with respect to the bulk, 
broadening the anomalous regions with a shift of the
characteristic $T$, $\rho$ and pressure $P$ towards
values higher than those in bulk. 
Furthermore, we find that in confinement the LLCP shifts towards $T$
higher than in bulk and that the $T$-increase is even stronger if we
calculate the phase diagram averaging over the entire fluid inside the
pore, including the heterogeneous part.  

The paper is organized as follows. In Section \ref{sec:methods} we 
describe the model and the simulation method.
In Section \ref{sec:results} we show our results about the phase
diagram and the anomalies of the homogeneous fluid, comparing the
confined and the bulk case. In Section \ref{sec:discussion} we
calculate the phase diagram under confinement including the
heterogeneous part of the fluid. In Section \ref{sec:conclusions} we
discuss our results and give our conclusions.

%
%

\section{Model and Method}
\label{sec:methods}

\begin{figure}[t!]
\begin{center}
\includegraphics[clip=true,width=12cm]{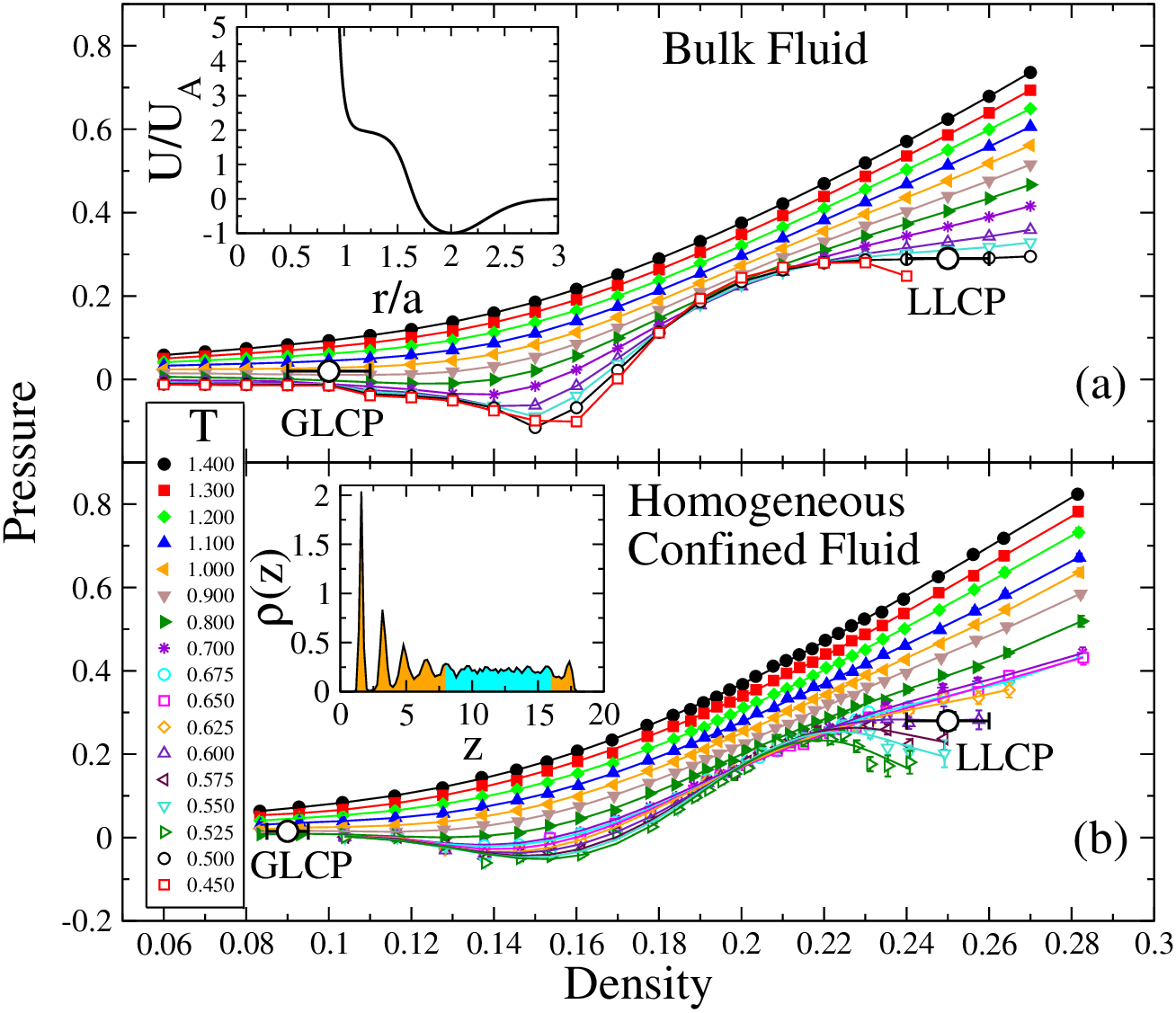}
\vspace{0.3cm}
\caption{\label{fig:3} 
Comparison between isotherms of the CSW interparticle
potential (Inset a) for bulk and confined fluids. (a) Isotherms in the
$P-\rho$ plane of the bulk fluid and (b) in the $P_{\|}$-$\rho$ plane of
the homogeneous fluid confined in a slit pore. In both panels
isotherms are between $T=1.4$ (top) and $T=0.45$ (bottom). Open
symbols with error bars and labels GLCP and LLCP denote our
estimates for the gas-liquid and the liquid-liquid critical point,
respectively. Inset b: A typical density profile $\rho(z)$ with
heterogeneous fluid (orange) near the walls and homogeneous fluid
(cyan) in the center of the slit pore. The example is at $T=0.7$ and
$\rho=0.22$. Isotherms in (b) are calculated by averaging only
over the homogeneous fluid, excluding the contribution coming from the
heterogeneous part.}   
\end{center}
\end{figure}
We consider the anomalous fluid described by the CSW effective
potential (inset Fig.~\ref{fig:3}a) defined as
\cite{Fr07a,OFNB08,vilaseca:084507,leoni2014}  
\begin{equation}
U(r)\equiv\dfrac{U_R}{1+\exp(\Delta(r-R_R)/a)}-U_A\exp\left[-\dfrac{(r-R_A)^2}
  {2\delta_A^2}\right]+U_A\left(\dfrac{a}{r}\right)^{24}   
\end{equation}
where $a$ is the diameter of the particles, $R_A$ and $R_R$ are the
distance of the attractive minimum and the repulsive radius,
respectively, $U_A$ and $U_R$ are the energies of the attractive well
and the repulsive shoulder, respectively, $\delta_A^2$ is the variance
of the Gaussian centered in $R_A$ and $\Delta$ is the parameter which
controls the slope between the shoulder and the well at $R_R$.
We choose the parameters $U_R/U_A=2$, $R_R/a=1.6$, $R_A/a=2$,
$(\delta_A/a)^2=0.1$, as in
Ref.s~\cite{OFNB08,Fr07a,vilaseca:084507,leoni2014}, and $\Delta=15$
to better emphasize the anomalies in density, diffusion and structure.
In order to reduce the computational cost, we impose a cutoff for the
interaction potential at a distance $r_c/a=3$ \cite{leoni2014}.

The particles are confined between two parallel walls separated along
the $z$ axis by a distance $L_z/a\simeq 19$. The distance is large
enough to have homogeneous fluid inside the pore \cite{leoni2014} for
a range of $T$ and $\rho$ comparable to that investigated for the bulk
case in previous works \cite{Fr07a,OFNB08,vilaseca:084507}.
The attractive wall is made of CSW particles quenched in a triangular
lattice with constant $d=a$ and centered at $z_{\rm attr}=0$.
The repulsive smooth wall has a repulsive interaction $U_{\rm
  rep}(z)\equiv[a/(L_z-z)]^9$ with the CSW particles at position 
$z$ \cite{varnik2000,leoni2014}.

We perform molecular dynamics (MD) simulations in the $NVT$ ensemble,
with $N=1024$ CSW particles in a constant volume $V$. We control $T$  
with the Allen thermostat  \cite{allen1989} and use the velocity Verlet
method \cite{allen1989} with time-step $dt=0.0032$ to integrate the
equations of motion.

Due to the chosen interactions with the walls, not all the volume $V$
is available to the CSW particles. 
Following Ref.~\cite{varnik2000}, the effective volume accessible to
particles is $V_{\rm eff}\equiv L_z^{\rm eff}A$, where $A\equiv
L_xL_y$ is the section of the simulation box and $L_z^{\rm eff}\simeq
L_z-a/2-(1/T)^{1/9}$ is the effective distance between the plates.
The effective density is $\rho_{\rm eff}\equiv \rho_{\rm
  eff}(\rho,T)\equiv \rho\cdot(L_z/L_z^{\rm eff})$. 
To explore different densities we change $L_x$ and $L_y$ so that we
can keep $L_z$ and $N$ constant to exclude finite-size effects when we
compare results for different $\rho$.
We adopt periodic boundary conditions in the $x$ and $y$ directions
parallel to the walls.

In slit pore geometries the isotropic bulk pressure $P$ is replaced by 
the pressure $P_{\|}$ parallel to the walls
\cite{klapp2007,Truskett01}, with $P_{\|}\equiv P_{\|}(z)\equiv
P_{xx}(z)\equiv P_{yy}(z)$, where $P_{xx}$ and $P_{yy}$
are the two diagonal components of the pressure tensor along the walls.
The third diagonal component gives the normal pressure $P_{\perp}=P_{zz}$,
while the non-diagonal components are zero (no shear forces are
present in the fluid) \cite{varnik2000}.   
The normal pressure $P_{\perp}$ can be computed as the net force along
the $z$ axis per unit area acting on one of the walls and at
(mechanical) equilibrium we verify that $P_{\perp}$ does not depend on
$z$, with $P_{\perp}\geq 0$ for all densities \cite{leoni2014}.
We calculate $P_{\|}$ using the virial expression for the $x$ and $y$
directions \cite{allen1989}  
\begin{equation}\label{equ:virial}
P_{\|}\equiv k_BT\rho_{\rm
  eff}-\dfrac{1}{2V}\left\langle\sum_i\sum_{j>i}\dfrac{x_{ij}^2+y_{ij}^2}{r_{ij}}\left(\dfrac{\partial  
U(r)}{\partial r}\right)_{r=r_{ij}}\right\rangle 
\end{equation}
where $k_B$ is the Boltzmann constant,
$r_{ij}\equiv |\vec{r}_i-\vec{r}_j|$ and $\langle \cdot \rangle$
represents  the thermodynamic average.

For the bulk case, after verifying that there are no relevant size
effects changing the value of $N$ from $1372$ to $676$, we simulate
$N=676$. This value is closer to the typical number of particles in
the homogeneous confined fluid (inset in Fig.~\ref{fig:3}b) minimizing
the relative size effect when we compare confined and bulk systems.
The bulk case with $N=1372$ has been extensively studied in
Ref.\cite{vilaseca:084507}.

We average the thermodynamic quantities over the homogeneous confined
fluid, after checking the stability of the system.
The homogeneous fluid region is defined as the part of the fluid for
which the following conditions are fulfilled (see
Ref.\cite{leoni2014}): (i) The density profile is constant (apart
from fluctuations consistent with the equilibrium compressibility)
with no sign of layering (absence of local maxima and minima in the
density profile).
(ii) In any plane parallel to the slit pore the homogeneous fluid
has a structurless radial distribution function, resembling a
two-dimensions section of a regular
fluid.
(iii) The characteristic decay time $\tau_{max}$ of the survival
probability (SP) of a particle in a finite section parallel to the
slit pore does not depend on the position of the section 
 within the homogeneous fluid.
(iv) The mean square displacement (MSD) of a particle within a
finite-width section parallel to the slit pore has a diffusive regime
independent on the position of the section  within the homogeneous
fluid.

We equilibrate the system by annealing from $T^*=4$ (although we
show results only for $T^*\leq 1.4$), where $T^*$ is the temperature
expressed in internal units (see below). For each temperature, the
system is equilibrated during $10^6$ time steps. We observe
equilibrium (no drift in average energy and 
pressure) after $10^4$ time steps for the range of $\rho$
and $T$ considered here \cite{leoni2014}.

All our results are averaged over 10 independent samples for the
confined fluid, and over 4 independent samples for the bulk.
Pressure, temperature, density, diffusion constant and time are all
expressed in internal units: $P^*\equiv Pa^3/U_A$, $T^*\equiv k_BT/U_A$,
$\rho^*\equiv\rho a^3$, $D^*\equiv D(m/a^2U_A)^{1/2}$, and $t^*\equiv
t(U_A/a^2m)^{1/2}$ respectively, with $m$ unit of mass. For sake of
simplicity, in the following we  drop the $^*$ and 
use the symbols $P$ and $\rho$ for the pressure and density, respectively, of
both bulk and confined fluids, dropping the labels $\|$ and 
``eff'' if not necessary, especially when we refer to both systems
at the same time.


\section{Results}
\label{sec:results}

\subsection{Comparison of the confined and bulk phase diagrams:
  liquid-liquid critical point and density anomaly.}

\begin{figure}
\begin{center}
\includegraphics[clip=true,width=12cm]{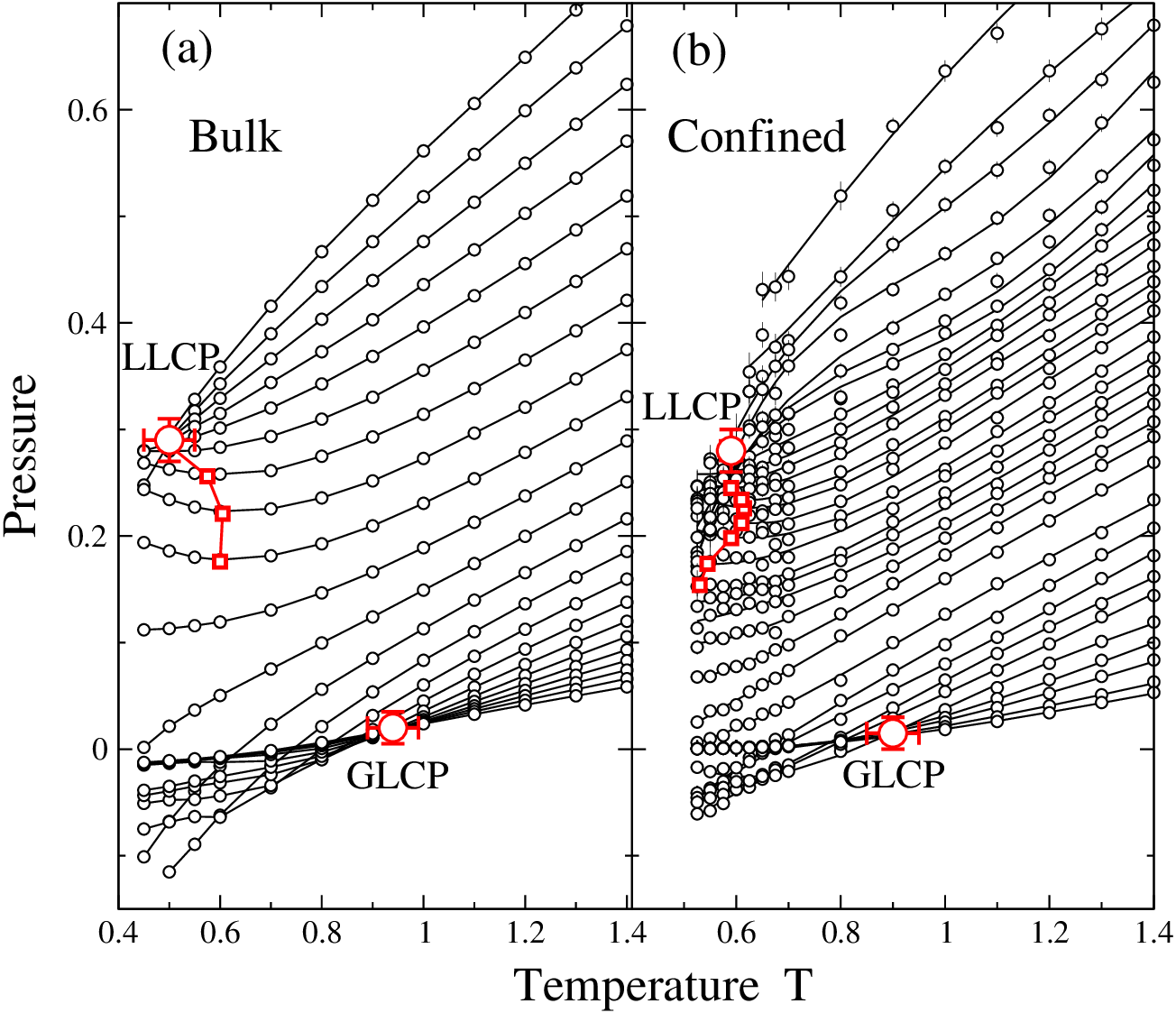}
\vspace{0.3cm}
\caption{\label{fig:4} Isochores in the pressure-temperature phase
  diagram projection for the CSW liquid. The line of temperature of
  maximum density (TMD) corresponds to the minima (red squares) along
  isochores (circles connected by lines) in the $P-T$ plane for bulk
  (a), and in the $P_{\|}-T$ plane for the confined fluid
  (b). In panel (a) the top isochore is for $\rho=0.27$ and the bottom
  for $\rho=0.06$ with $\Delta\rho=0.01$ density difference from one
  to the next. In panel (b) the top isochore is for $\rho\simeq 0.281$
  and the bottom for $\rho\simeq 0.083$ with $\Delta\rho\simeq 0.01$
  density difference. In both panels GLCP and LLCP (large red circles)
  mark the gas-liquid and liquid-liquid critical points, respectively.}  
\end{center}
\end{figure}

We calculate the phase diagram for the homogeneous confined fluid and
for the bulk fluid finding in both cases the GLCP and the LLCP 
(Figs.~\ref{fig:3} and \ref{fig:4}).
Following the standard definition, we estimate the critical point
parameters (Table~\ref{Tab}) by calculating  
$\partial P/\partial \rho|_T= 0$ and
$\partial^2 P/\partial \rho^2|_T= 0$ at high $T$ and low
densities (for GLCP) or high densities (for LLCP)
\cite{OFNB08,vilaseca:084507}\footnote{We observe (Fig.~\ref{fig:3})
  that the fluid phase for both bulk and confined cases corresponds to
  a density range smaller with respect to that of a normal fluid, as
  for example the Lennard-Jones (LJ).  
Indeed, one of the common features of the potentials for
anomalous-liquids is the very low density of the liquid phase, due to
the characteristic soft-core interaction. For example, the LJ
potential has its minimum at interparticle distance
$r_{min}^{LJ}/a=2^{1/6}$, while the CSW has its  minimum at
$r_{min}^{CSW}/a\simeq 2$, where $a$ is the particle diameter. Hence,
for typical liquid configurations of the two systems it is
$r_{min}^{CSW}/r_{min}^{LJ}\simeq 2/2^{1/6}\simeq 1.78$, or
equivalently $\rho^{CSW}/\rho^{LJ}\simeq
(r_{min}^{LJ}/r_{min}^{CSW})^3\simeq 0.18$.}\footnote{By definition,
  the $\rho$-derivative of the critical isotherm is 
  zero. From the other hand, the point where minimia and maxima showen
  by isotherms meet (corresponding to $\partial^2 P/\partial
  \rho^2|_T\rightarrow 0$), identify the critical point.}.

The comparison of the critical values for the confined fluid  and for
the bulk fluid (Table~\ref{Tab}) reveals that the confinement changes
the critical temperature $T_{\rm LLCP}$ of
LLCP. In particular, $T_{\rm LLCP}$ for the confined case is $(19\pm 6)$\%
larger than the bulk value.

For both bulk and confined fluid we find minima along isochores and
we identify the line of temperatures of maxima density (TMD) by
calculating numerically $\partial P/\partial T|_{\rho}=0$ and
$\partial^2 P/\partial T^2|_{\rho}>0$ (Fig.~\ref{fig:4}). As seen for
the bulk \cite{OFNB08,vilaseca:084507}, we observe that the TMD line
converges toward the LLCP in the $P$-$T$ plane.

\begin{table}[]
  \centering
  \begin{tabular}{| r | c | c | c | c | c | c |}
    \hline
CP    & Density GL
    &  Pressure GL
    & Temp. GL
    & Density LL
    &  Pressure LL
    & Temp. LL
    \\
    \hline
    \hline
     Bulk & 0.100$\pm$0.010 & 0.020$\pm$0.015 & 0.94$\pm$0.05 &
     0.25$\pm$0.01& 0.29$\pm$0.02& 0.50$\pm$0.05\\
   Confined &  0.090$\pm$0.005 & 0.015$\pm$0.015 & 0.90$\pm$0.05 &
   0.25$\pm$0.01& 0.28$\pm$0.02& 0.59$\pm$0.02\\ 
    \hline
  \end{tabular}
  \caption{\label{Tab}Estimates of the parameters for the gas-liquid critical
    point GLCP and the liquid-liquid critical point LLCP for the
    bulk fluid (upper row) and the homogeneous confined fluid (lower row).}
  \label{T1}
\end{table}

It is interesting to ask if the knowledge of the phase diagram of the
confined fluid can give us information about the bulk system.
To this goal we look for a simple global transformation that
could fairly rescale the entire set of bulk on the confined isotherms.
In particular, we test if such a rescale is possible by
applying a transformations $\rho_{tr}\equiv a\rho+b$ and
$P_{tr}\equiv cP+d$ to the bulk quantities, where the coefficients
$a$, $b$, $c$ and $d$ depend linearly on the temperature as
$a(T)\equiv a_0T+a_1$ with $a_0$  and $a_1$ constant and so on for
$b$, $c$ and $d$.
To optimize the rescaling we minimize the cumulative deviation 
\begin{equation}
\delta\equiv \sum_i\delta_i\equiv \sum_i\left[\int_{\rho_{\rm
      min}}^{\rho_{\rm
      max}}(P_{tr}(T_i)-P_{\|}(T_i))^2d\rho\right]^{-1/2},
\label{delta}
\end{equation}
where $T_i$ refers to the temperature of the $i$-th isotherm and
$[\rho_{\rm min}, \rho_{\rm max}]$ is the density range of comparison. 

\begin{figure}
\begin{center}
\includegraphics[clip=true,width=12cm]{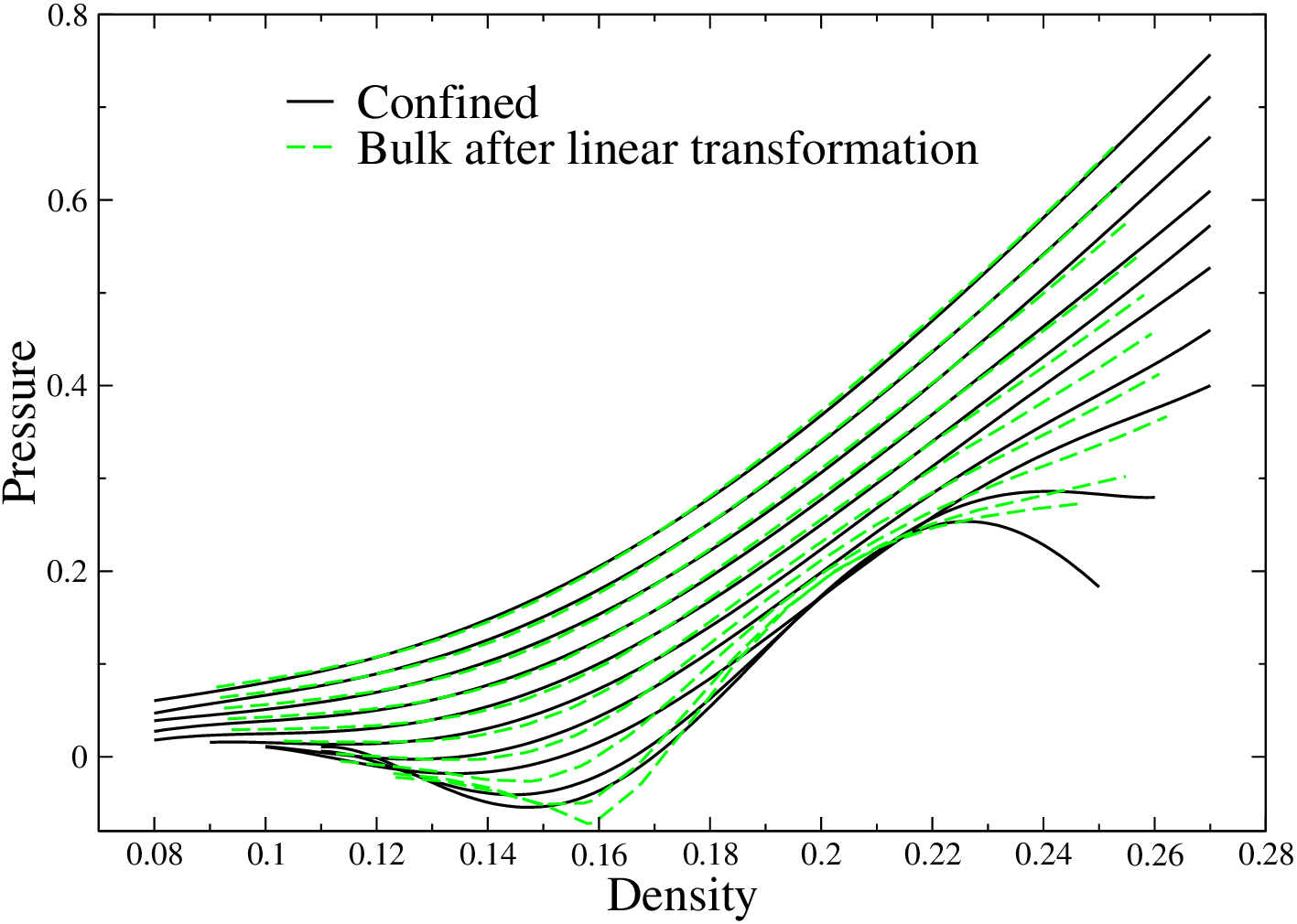}
\vspace{0.3cm}
\caption{\label{fig:5} A global linear transformation allows us to
  compare isotherms of the bulk and the confined CSW
  fluids. Isotherms of the confined homogeneous fluid (continuous
  black lines) and the bulk fluid after the global transformation
  (dashed green lines) are from $T=1.4$ (top) to $0.55$ (bottom). 
  The deviation $\delta$, in Eq.(\ref{delta}), between the two sets of
  isotherms  is minimized for the following set of coefficients:
  $a_0=-0.02$, $a_1=0.912$, $b_0=0.00$, $b_1=0.02$, $c_0=-0.02$,
  $c_1=1.12$, $d_0=-0.02$, $d_1=0.02$.}
\end{center}
\end{figure}

We find (Fig.~\ref{fig:5}) that there is a linear
transformation that fairly superimposes the two set of isotherms at
high temperatures, while at low $T$ we observe deviations. 
In particular, for $T>T_{LLCP}$ the two set of isotherms are
qualitatively and quantitatively closely related, while for
$T<T_{LLCP}$ the deviations are strong at high densities but there is
still a qualitative agreement at intermediate densities corresponding
to the low density liquid phase.

%
%
\subsection{Structural anomaly}

The translational order in a fluid can be
estimated by the quantity \cite{Sh02,Er01,Er03} 
\begin{equation}
  t \equiv \int_0^\infty |g(\xi)-1|d\xi
  \label{t}
\end{equation}
where $\xi\equiv r\rho^{1/3}$ is the distance $r$ in units of the mean
interparticle separation $\rho^{-1/3}$ and $g(\xi)$ is the radial
distribution function.
For an ideal gas $g(\xi)=1$ at any distance, hence $t=0$ corresponding
to the absence of translational order. For a normal fluid at constant
$T$ and increasing $\rho$, the deviation of  $g(\xi)$ from 1 increases
monotonically, implying a monotonic increase of $t$.

For the CSW bulk fluid at low enough $T$ there is a range of $\rho$
within which the order parameter $t$ decreases for increasing $\rho$
along isotherms \cite{OFNB08,vilaseca:084507}. 
This anomalous behavior indicates  that the translational order is
progressively reduced for increasing $\rho$. 
At high enough $\rho$, the CSW bulk fluid recovers the normal behavior
with $t$ increasing for increasing $\rho$ \cite{Vilaseca2011,Er01}.

Here we investigate how the confinement affects this property.
For the homogeneous fluid sub-region in the confined system we
calculate $t$ as in Eq.~(\ref{t}) integrated up to a cutoff distance
$\xi_c=L_z^{\rm hom}\rho^{1/3}/2$, where $L_z^{\rm hom}$ is the thickness
of the homogeneous sub-region, because we observe that
$g(\xi)\simeq 1$ for $\xi>\xi_c$. 

\begin{figure}[t!]
\begin{center}
\includegraphics[clip=true,width=12cm]{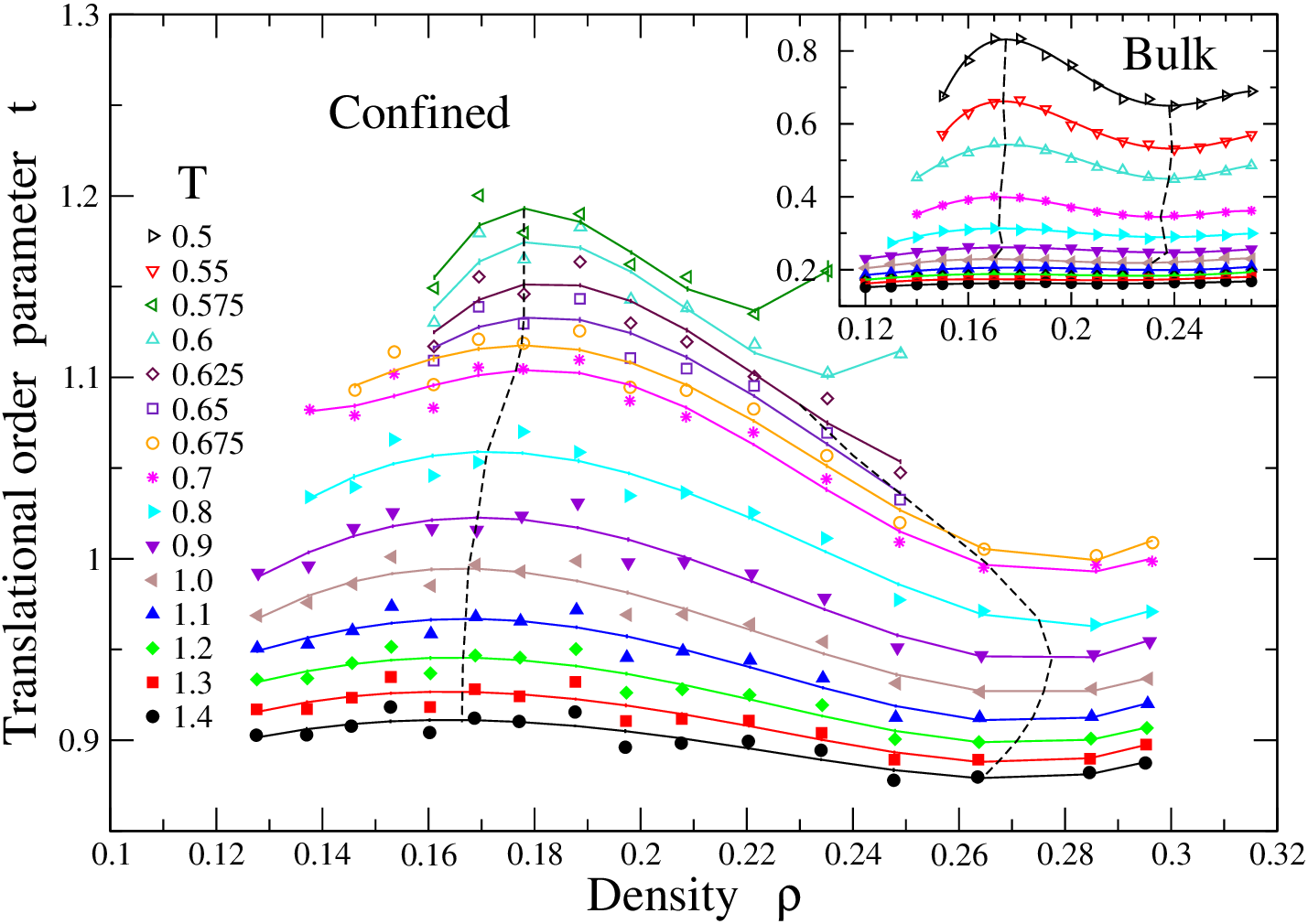}
\vspace{0.3cm}
\caption{\label{fig:6} Translational order parameter $t$ for the
  homogeneous CSW fluid confined in a slit pore from $T=0.575$ (top) to
  $T=1.4$ (bottom). Inset: 
  $t$ for the CSW bulk fluid from $T=0.5$ (top) to
  $T=1.4$ (bottom). In both panels dotted lines are guides for the
  eyes delimiting the anomalous region where t decreases for
  increasing density.}
\end{center}
\end{figure}

We find that $t(\rho)$ has maxima and minima for $T\leq 1.4$,
delimiting an anomalous region (Fig.~\ref{fig:6}).  
We observe that for the confined CSW fluid the anomaly in $t$ extends at
$T$ and $\rho$ that are larger than for the bulk CSW fluid
(Fig.~\ref{fig:6}, inset).

A striking difference is that in confinement $t$ is much larger than
in bulk.
We find that the difference increases for increasing $T$ and
increasing $\rho$. For example, at $T=0.6$ for the confined case 
$t$ is $\simeq 114$\% larger than for the bulk at low $\rho$ and
$\simeq 145$\% larger  at high $\rho$, while at $T=1.4$ is
$\simeq 456$\% larger at low $\rho$ and $\simeq 483$\% larger at high
$\rho$. 
Therefore, the confinement largely increases the translational order
of the fluid and the change is stronger at higher $T$ and $\rho$.

%
%
\subsection{Diffusion anomaly}

In normal fluids, the diffusion coefficient monotonically decreases
for increasing density along isotherms.
The CSW bulk fluid is anomalous also in its diffusion behavior.
At low enough $T$ there is a density interval within which the
diffusion coefficient $D_{\rm bulk}$ increases for increasing density
\cite{OFNB08,vilaseca:084507}.  

To study the diffusion properties of the confined CSW fluid and compare
with the bulk, we make the following observations.
To correctly define  the diffusion constant in the confined case it is
necessary (i) to reach the diffusion regime for the mean square
displacement (MSD) and (ii) to be sure that the particles explore an
environment that is homogeneous over the diffusion time-scale,
otherwise the heterogeneities that survive over large time-scales
could lead to apparent superdiffusive or subdiffusive regimes
\cite{leoni2014}. 
Condition (i) can be satisfied thanks to our set-up with periodic
boundary conditions parallel to the walls, while to guarantee
condition (ii) we focus on those particles that moves only within the
region of the slit pore where the liquid phase is homogeneous (Inset
Fig.~\ref{fig:3}b). 

For the particles in the homogeneous fluid we define the MSD parallel
to the walls as \cite{leoni2014} 
\begin{equation}
\langle(\Delta
r_{\|}(\tau))^2\rangle\equiv\langle(r_{\|}(t-t_0)-r_{\|}(t_0))^2\rangle, 
\end{equation}
where $r_{\|}\equiv (x^2+y^2)^{1/2}$ and $\tau\equiv t-t_0$ is the
time spent in the homogeneous region of the pore by a particle that
stays or enters in this region at time $t_0$.
The diffusion coefficient $D_{\|}$ in the direction parallel to the
walls is by definition 
\begin{equation}
D_{\|}\equiv \lim_{\tau\rightarrow\infty}\langle(\Delta
r_{\|}(\tau))^2\rangle/(4\tau). 
\end{equation}

\begin{figure}
\begin{center}
\includegraphics[clip=true,width=8cm]{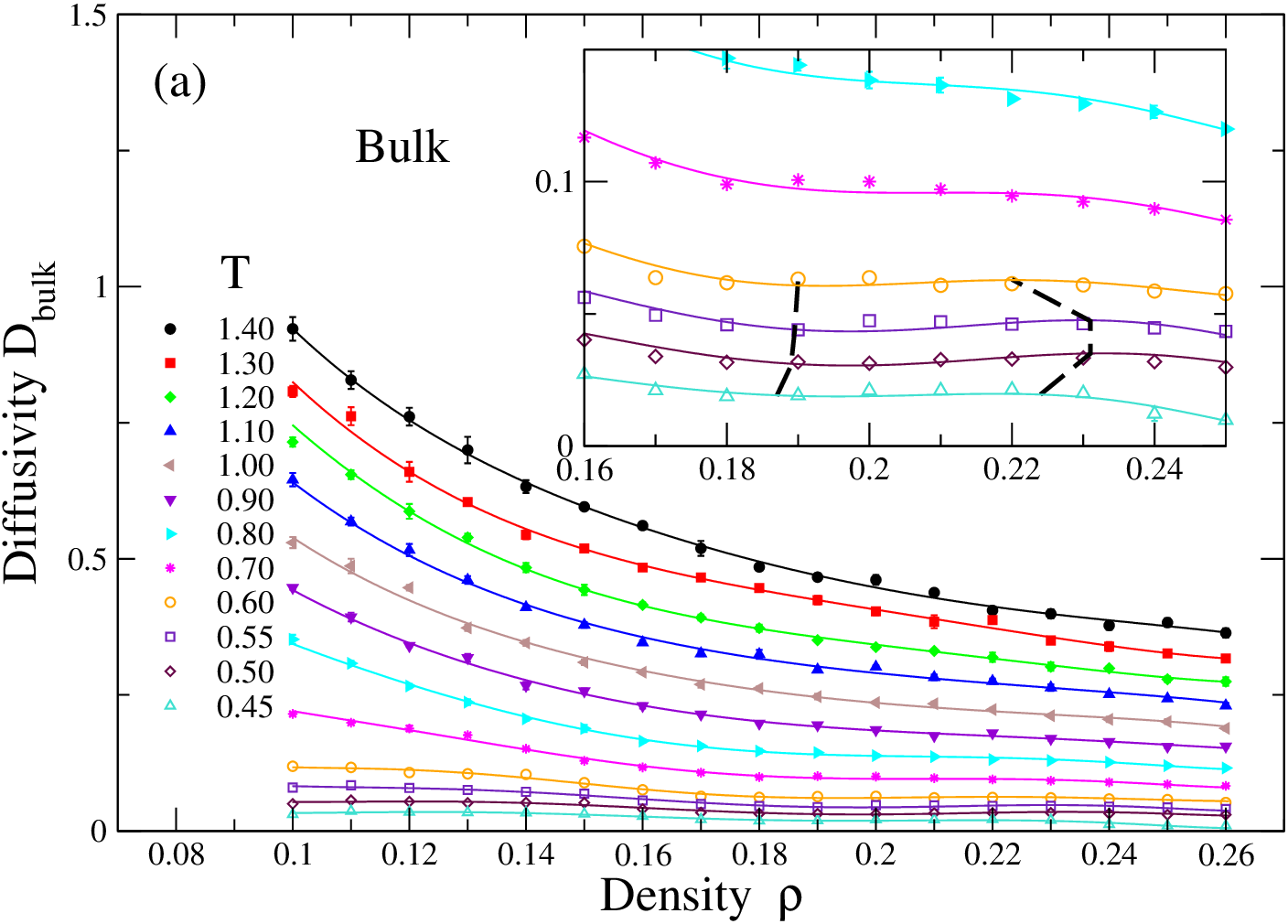}
\includegraphics[clip=true,width=8cm]{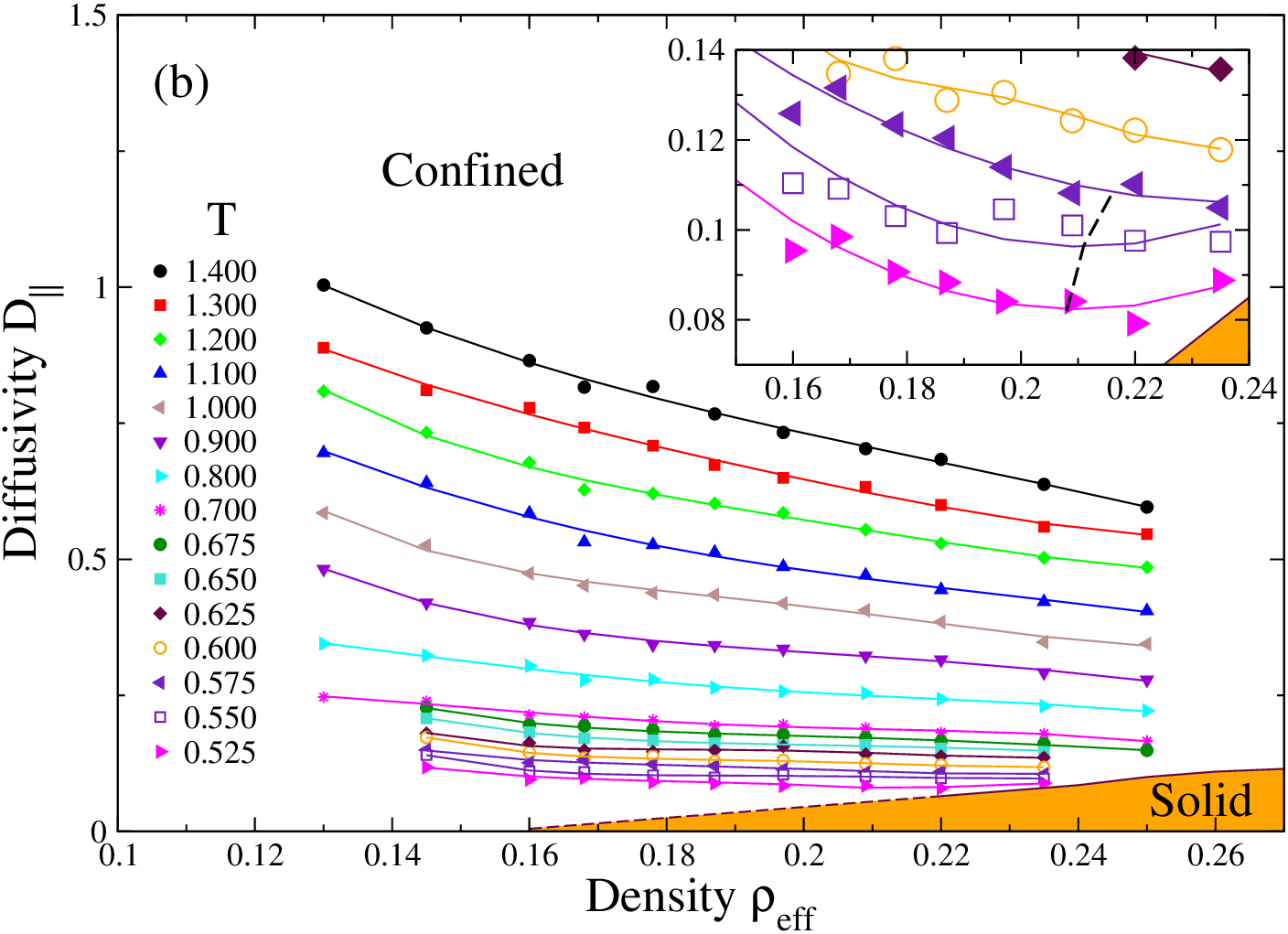}
\vspace{0.3cm}
\caption{\label{fig:78} Anomalous behavior of the constant-$T$
  diffusion coefficient  for the bulk and confined CSW fluid.
  (a) For the bulk we calculate 
  $D_{\rm bulk}$ between $T=1.40$ (top) and
  $T=0.45$  (bottom) and find 
  a density of minimum diffusion
  $D_{\rm  bulk}^{\rm min}$ and a density of maximum diffusion
  $D_{\rm  bulk}^{\rm max}$ (both marked by dashed lines) for 
    $T\leq 0.55$ (inset).
  (b) For the homogeneous confined fluid we calculate 
    $D_{\|}$ between $T=1.40$ (top) and 
  $T=0.525$ (bottom) and  find an anomalous behavior, with  
  a density (dashed line) of minimum diffusion
  $D_{\rm  \|}^{\rm min}$
  for  $T< 0.60$ and $\rho>0.21$ (inset).
  The region with anomalous 
  diffusion is limited at higher densities by the
  heterogeneous 
  crystallization of the fluid under confinement  
    (orange area).}
\end{center}
\end{figure}

Comparing the calculations for $D_{\rm bulk}$ and $D_{\|}$
(Fig.~\ref{fig:78}) we observe that the diffusion anomaly in the
confined system starts at higher densities with respect to the bulk
case. In particular, for the bulk we find a normal behavior at high
$T$ and the diffusion anomaly at low $T$ marked by a density of
minimum diffusion $D_{\rm  bulk}^{\rm min}$ at $\rho\simeq 0.19$ 
and a density of maximum diffusion $D_{\rm  bulk}^{\rm max}$ at
$\rho\simeq 0.23$ along isotherms with $T\leq 0.60$. 

On the other hand, for the confined fluid we find anomalous diffusion
at low $T$ in a range of density that starts at the density of minimum
diffusion $D_{\rm\|}^{\rm min}$ at $\rho\simeq 0.21$ along isotherms
with $T< 0.60$. However, due the heterogeneous crystallization under
confinement occurring at higher $\rho$ \cite{leoni2014}, we cannot
observe the high-density limit of the anomalous diffusion region
(Fig.~\ref{fig:78}). 

We find that $D_\|$ under confinement is much larger than
$D_{\rm  bulk}$  and that the difference increases for
decreasing $T$ and increasing $\rho$. For example, at $T=1.4$ 
is $D_\|\simeq 150\%~ D_{\rm  bulk}$ at low $\rho$ and
$D_\|\simeq 162\%~ D_{\rm  bulk}$ at high $\rho$, while at
$T=0.55$ is $D_\|\simeq 205\%~ D_{\rm  bulk}$ at low $\rho$ and
$D_\|\simeq 206\%~ D_{\rm  bulk}$ at high $\rho$.
Therefore, the confinement largely enhances the translational dynamics
of the homogeneous fluid with respect to the bulk and the change is
stronger at low $T$ and high $\rho$.

\subsection{Hierarchy of anomalies} 

\begin{figure}
\begin{center}
\includegraphics[clip=true,width=8.cm]{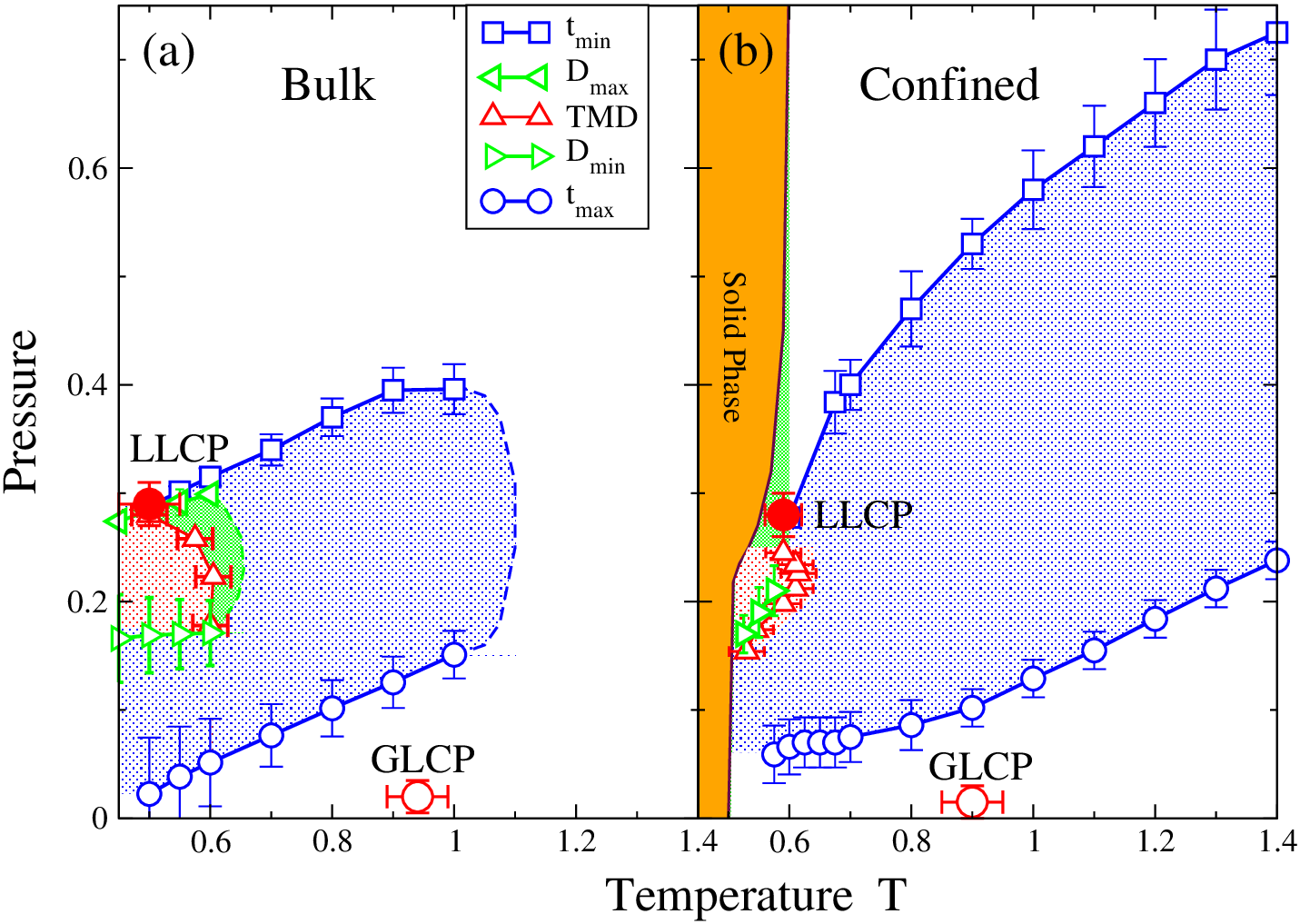}
\includegraphics[clip=true,width=8.cm]{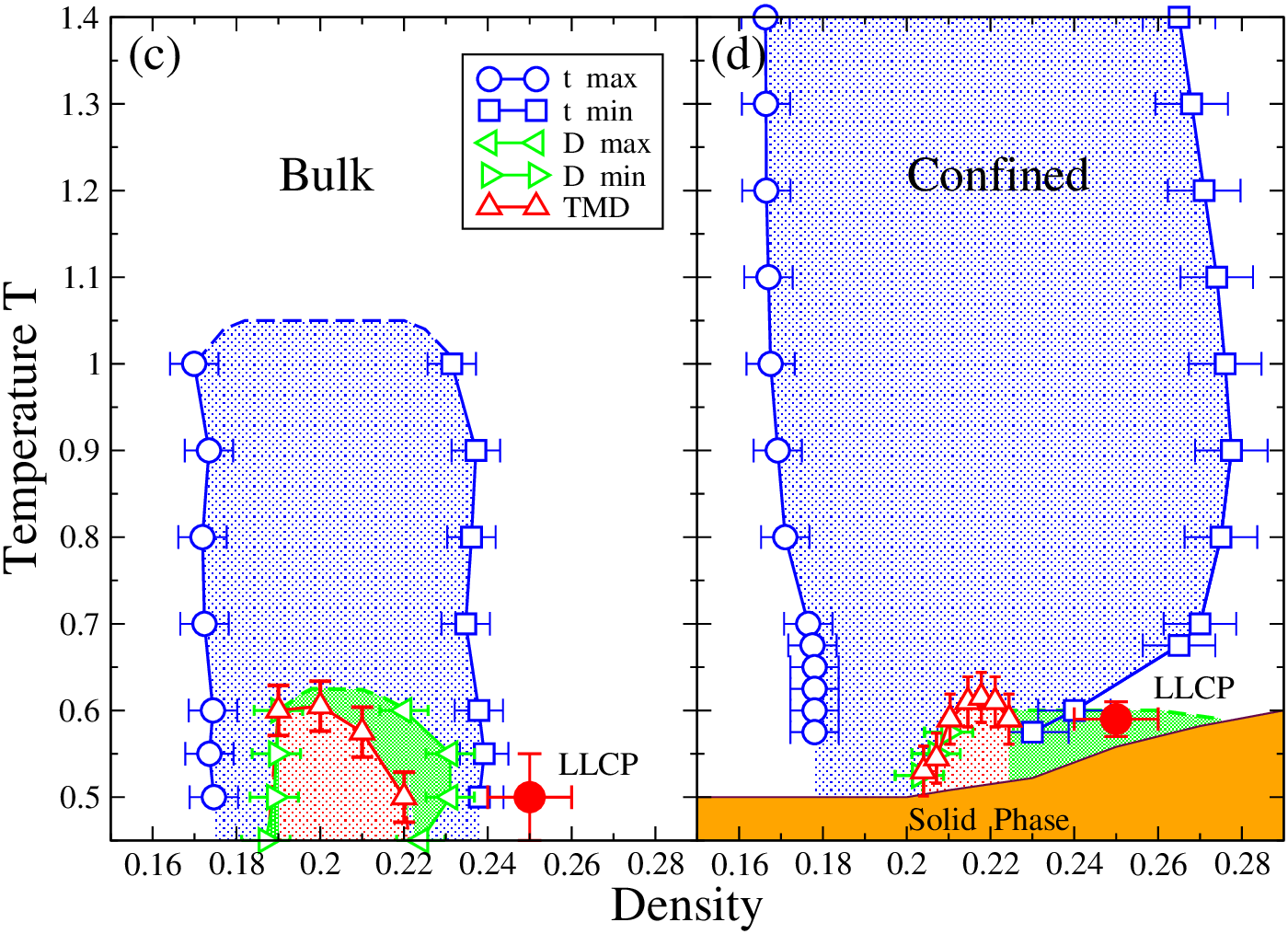}
\vspace{0.3cm}
\caption{\label{fig:9} Water-like hierarchy of anomalies,
  in the pressure-temperature (a,b) and temperature-density (c,d) planes,
  for the bulk (a,c) and the homogeneous confined fluid (b,d). We find
  the GLCP (open red circle) and
  the typical nested domes of structural-anomaly (blue) region,
  delimited by the loci of  $t_{\rm max}$  (blue circles) and $t_{\rm
    min}$ (blue squares),  the diffusion-anomaly (green) region, and
  the thermodynamic-anomaly (red) region marked by the TMD line (upper
  red triangles). For the bulk (a,c) the diffusion-anomaly region is
  delimited by the loci of $D_{\rm min}$ (green right triangles) and
$D_{\rm max}$ (green left triangles), while for the confined fluid (b,d)
the region starts at higher density with respect to the bulk case and
is limited by the nucleation of the solid  phase (orange region)
that prevents us from observing the locus of $D_{\rm max}$.
In both systems the LLCP (solid red circle) occurs near the
high-pressure, high-density limit of the anomalous regions.}
\end{center}
\end{figure}

The CSW particles in bulk display a water-like hierarchy of anomalies
\cite{OFNB08,vilaseca:084507}.  Here we find the same hierarchy also
for the confined CSW fluid (Fig.~\ref{fig:9}).
In particular, for both systems  we find that the
structural-anomaly region, between  the loci of $t_{\rm max}$
and $t_{\rm min}$, includes the diffusion-anomaly region. The latter,
in turn,  encloses the density-anomaly region delimited by the TMD line.
This hierarchy of anomalies is similar to that observed in other
models with water-like properties \cite{vilaseca:084507} and differs
from the one characteristic of silica-like fluids. In silica  the
diffusion-anomaly region encloses the region of structural
anomalies, which contains the TMD line, marking the region of
thermodynamic anomalies \cite{Shell2002}. The hierarchy also differs
from those seen in other isotropic model potentials with water-like
anomalies \cite{Xu06,Fomin2011}.

Despite the hierarchy is unchanged, our calculations reveal that the
confinement has a strong effect on the regions of the anomalies
when compared with the results for the bulk fluid (Fig.~\ref{fig:9}).
The more evident difference between the bulk and the confined case is the
extension of the structural-anomaly region.
In bulk we observe the structural anomaly only for $T<1.05$, while for
the confined fluid we find the anomaly for all the temperatures we
investigate, $T\leq 1.4$. Furthermore, the translational order $t$ for
the confined fluid is anomalous at densities (and pressures) larger
than those for the bulk case. It is interesting to observe that the
high-density limit of the structural anomaly for the confined case
seems to be strongly correlated at low $T$ to the filling up of the
pore with the solid phase (Fig.~\ref{fig:9}). This observation
suggests that the anomalous reduction of the translational order for
increasing density could be the leading cause for the suppression of
the crystal phase in confinement. Once the translational order
recovers a normal behavior, at density higher than the locus of 
$t_{\rm min}$, the fluid starts to crystallize.

Another difference between the bulk and the confined system is the
possibility to explore the entire region of diffusion anomaly. While in
bulk we observe both loci $D_{\rm min}$ and $D_{\rm max}$, under
confinement we can find only the first, marking the onset of the
diffusion-anomaly region, but not the second, marking the high-density
(high-pressure) limit of the same region. This is because at high
density and high pressure the slit pore is filled with  heterogeneous
fluid coexisting with the crystal phase, leaving no space for the
homogeneous fluid. Hence the diffusion-anomaly region for the confined
fluid, although extending over a wider range of densities (and
pressures) with respect to the bulk case, is limited by the crystal
phase. 

Regarding the thermodynamic anomaly we observe that  the TMD line in
the confined fluid starts at higher density with respect to the
bulk. The TMD extends on a smaller range of densities with no relevant
change in $T$ compared with the bulk.

%
%

\section{Discussion}
\label{sec:discussion}

\subsection{Confinement effects on structural and thermodynamic anomalies}

Our results show that the confinement induces a shift of the anomalous
regions and in particular of the largest, i.e. the structural anomaly region, 
towards larger densities, pressures and temperatures with respect
to the bulk. This effect is commonly observed in confined fluids 
regardless the strength of attraction of the wall and its structure 
\cite{Krott2015,sun2015}.
It can be understood as a result of the large local increase of
density, due to the layering parallel to the walls, and of the
consequent increase of particle-particle repulsion and overall
pressure. 

On the other hand, the effect of the confinement on the maximum
temperature of the anomalies is less intuitive.
Previous studies about core-softened fluids under confinement have
shown that density and diffusion anomalies shift  towards
lower $T$ regardless of the surface structure
\cite{Krott2015} or, for the TMD, regardless of the surface attraction
\cite{sun2015}. Here we observe that the TMD is not affected 
in $T$, while the  diffusion-anomaly region has a small decrease in
$T$, consistent with Ref. \cite{Krott2015}. On the contrary, the
effect of confinement on the structural-anomaly region, for which
there are no data in Ref.s \cite{Krott2015, sun2015}, is a large
increase in $T$. We interpret this result as a consequence of the
presence of the attractive wall. 
The attractive interaction with the structured wall 
induces an order that propagates through the layers \cite{leoni2014}
largely increasing the translational order parameter $t$. This
effect is stronger at high $T$ and $\rho$ extending
the anomalous structural region toward high $T$ and $\rho$.

\subsection{Confinement effects on dynamics}

We observe that also the dynamics of the fluid is strongly affected by
the confinement. As reported in Ref.~\cite{leoni2014}, layering
starting near the attractive wall induces dynamics heterogeneities
that disappear only within the central homogeneous part of the
confined fluid. Here we find a  large enhancement of the diffusion
constant of the homogeneous fluid with respect to the bulk.

This effect poses questions when we compare our results to other
water-like models, atomistic water simulations and experiments in
hydrophobic (e.g., carbon-based) and hydrophilic (e.g., silica-based)
confinements. 
In particular, the enhancement is at variance with simulation results
for water in hard-core repulsive smooth slit pores with size $<7$
layers (we use the fact that each water layer is on average 0.3~nm
wide) \cite{Liu2004}, coarse-grained water in slit pores of width
3--7 layers \cite{Eslami2013}, water in disordered carbon slit
pores with a size distribution 1--3 layers range
\cite{Nguyen2012}, water in cylindrical hydrophobic and hydrophilic
pores of sizes $<3$ layers \cite{Allen1999}.
Furthermore, the structured attractive wall should decrease
the fluid mobility with respect to the smooth wall for isotropic
potentials \cite{Krekelberg2011,Krott2015}. Also, some experiments show
a dramatic increase of the mechanical relaxation times of water 
compressed between hydrophilic surfaces to less than 3--4 molecular
layers \cite{Khan2010}, or a mild increase, within a factor of three
of the bulk value, of the effective viscosity of water confined
between curved mica surfaces even when it is confined to a number of
layers $\leq 10$ layers \cite{Raviv2001}. 

Nevertheless, the slowdown of the dynamics under confinement is quite
controversial. For example, simulations of water confined between
parallel slabs of quartz with a separation of 3 to 8 water layers
indicate a lateral diffusion coefficient comparable to the bulk
\cite{Zangi2004}. 
Other simulations show that water in carbon nanotubes with $\lesssim
3$ layers diameter undergoes normal diffusion, instead of 
subdiffusion, as a consequence of strong correlations due to hydrogen
bonding between neighboring water molecules  \cite{Mukherjee2007}, or
has an enhanced diffusion when anisotropy in the C-H and C-O
interactions are considered \cite{Perez-Hernandez2013}.
Also simulations in silica nanopores show that  the first statistical
water monolayer is immobile and the rest of the water is
free and behaves like bulk water as far as cooperative properties,
such as diffusion coefficient, are concerned for pores of 5 layers
\cite{Gallo2010} and of 3 to 13 layers diameters \cite{Bourg2012}.
In other cylindrical hydrophilic pores,  simulations show that
confined water has translational mobility enhanced relative to bulk
water in the core region of amorphous silica with diameters ranging
from 9 to 13 layers \cite{Milischuk2011} and in TiO2-rutile pores of
diameter $\gtrsim 9$ layers \cite{Solveyra2013}.

Several experiments with carbon nanotubes reported a water flow
exceeding values calculated from continuum hydrodynamics models by more
than 3 orders of magnitude for pore diameter $<7$ layers \cite{Holt2006}
or from 4 to 5 orders of magnitude for $\simeq 23$ layers pore diameter
\cite{Majumder05, Majumder2011}.
Ultrafast water flow was observed also in experiments and simulations
of layered structures of graphene-based membranes  separated by a
typical distance of $\simeq 3$ layers \cite{Nair2012}.
However, other experiments showed that the water flow rate through
individual (1~mm) ultralong carbon nanotubes  with diameters of 3--6
layers has an enhancement below 3 orders of magnitude \cite{Qin2011}.   

The experimental superfast water flow in carbon nanotubes has
been tested in simulations with contradictory results.
For example, first principle molecular dynamics simulations for water
confined in single-wall carbon nanotubes and graphene sheets 3--8 layers
apart showed \cite{cicero2008} only a moderate increase in the liquid
self-diffusion coefficient not consistent with the enhancement by
orders of magnitude found in Ref.~\cite{Holt2006}.
Also, it was found that transport enhancement rates depend on the
nanotube's length  and are only 2 orders of magnitude over
the continuum predictions for tubes with inner diameter of $\simeq 6$
layers \cite{Walther2013}. Nevertheless, more recent hybrid
molecular-continuum simulations showed that it could be possible to
reconcile all the simulations and experimental results by taking into
account all the assumptions made during both the calculations and the
experiments and the frictional properties for long nanotubes
\cite{Ritos2015}. 

Other simulations showed a non-monotonic behavior of diffusion
coefficient as function of the confinement size. For example,
it has been found a slight enhancement in the diffusion coefficient
of water confined in  carbon nanotubes of diameter $\gtrsim 7$ layers
compared to bulk water with a sharp decrease and a minimum at $\simeq
3$ layers and an increase at smaller diameters \cite{Ye2011,
  Barati-Farimani2011}, consistent with a non-monotonic  flow
enhancement observed in experiments for ultralong tubes with diameter
around $\simeq 3$ layers \cite{Qin2011}. 
Non-monotonicity in the diffusion has been reported also for a
core-softened fluid confined in nanotubes of different radiuses, with
a minimum diffusion for a radius approximately between 2 and 4 layers
\cite{Bordin2012}, and for confined films of spherical particles when
the separation between the attractive plates is around 4 and 5 layers
\cite{Gao1997}. 

In our case, the confinement induces a diffusion coefficient that is
at most twice as large as in bulk, hence inconsistent with ultrafast
water flow, but consistent with the moderate increase of mobility
found in several of the works cited above. For the range of densities
studied here, we have between 3 and 8 layers of homogeneous fluid
confined within heterogeneous layers of amorphous solid near the walls of 
the slit pore. This range of number of layers corresponds, as
mentioned above, to the pore sizes for which there are experiments and
simulations showing a moderate mobility increase or a non-monotonic
behavior of the dynamics. 
In particular, the largest increase of the diffusion occurs  at low
$T$ and high $\rho$, i.e. when there are only a few (3 or 4) layers of
homogeneous fluid confined within the slit pore. 

Our interpretation is that the observed layering induces collective
diffusion modes and a related higher mobility.  A similar mechanism
due to packing and layering has been proposed for room-temperature
ionic liquids in nanoporous media \cite{Huber2015}. Also in confined
water clustering, due to hydrogen-bond correlation, has been
considered as the cause for increased mobility
\cite{Perez-Hernandez2013}. 
In our case  layering \cite{leoni2014} and long-range correlations and
structural changes \cite{vilaseca:084507} can be the origin of
correlated aggregates and collective modes responsible for the
moderate increase of mobility. 

\subsection{Confinement effects on the LLCP}
 
While the slit pore does not change significantly the critical density
and pressure of the LLCP, it lifts the critical temperature to higher
values with respect to the bulk.
A similar effect was observed for water under hydrophilic,
i.e. attractive, confinement \cite{brovchenko2007}.
However, a number of isotropic models with water-like properties in
nano-confinement have shown that the LLCP shifts towards lower $T$
independently of the surface structure \cite{Krott2015} or the surface 
interaction \cite{sun2015}\footnote{These results somehow recall those
  for the case of wetting in a slit-pore confinement with 
competitive surface field, studied by means of the Ising model, 
for which theory \cite{parry1990} and
simulation \cite{albano2012} show that the critical temperature
in confinement is lower than in bulk. However, two-states models, such
as the Ising model, do not allow for the range of density fluctuations
that are important in our study}.

Nevertheless, we understand the shift of the LLCP to high $T$ as a
consequence of the
strength of the interaction between the fluid and the attractive wall
irrespective of the structured, smooth or amorphous nature of the
confinement \cite{Truskett01}.  Our result is consistent with 
Ref.~\cite{brovchenko2007} where the fluid-wall interaction is strong
enough to induce a high-density layer near the wall. Further support
to our interpretation comes from the observation that in those cases
in which the walls are repulsive \cite{Krott2015} or not enough
attractive to form a high-density layer near the  wall \cite{sun2015},
the shift of the LLCP towards low $T$ decreases for increasing density
of the layer near the wall. 

We observe that both in bulk and under confinement the LLCP occurs
just above the high-density (high-pressure) limit of the anomalous regions.
We understand this as due to the necessity for the fluid to first 
recover the normal (increasing) behavior of $t$ for increasing $\rho$
and then reach the value of $t$  large enough for the liquid-liquid
phase transition. Under confinement further increase of $\rho$ induces the
long-range translational order characteristic of the crystal phase and
the consequent crystallization. This interpretation is consistent with
the fact that the anomalous behavior of $t$ correlates
 with the limit of stability of the liquid with respect to the
crystal under confinement (Fig.~\ref{fig:9}d).

The fact that the LLCP occurs near the high-pressure, high-density
limit of the anomalous regions suggests that the knowledge of
the limits of these regions could give an indication about
where the LLCP is located. 
This finding together with the result that the LLCP for the confined
fluid is lifted up in temperature, above the limit of the solid-like
phase, gives support to the idea that the observation in an
appropriately defined confinement could be a way to make a LLCP more
accessible to the experiments.  In thie light of these observations,
in the following we discuss the possible implications of our results
for experiments. 

\subsection{Phase diagram of the entire fluid}

\begin{figure}
\begin{center}
\includegraphics[clip=true,width=12cm]{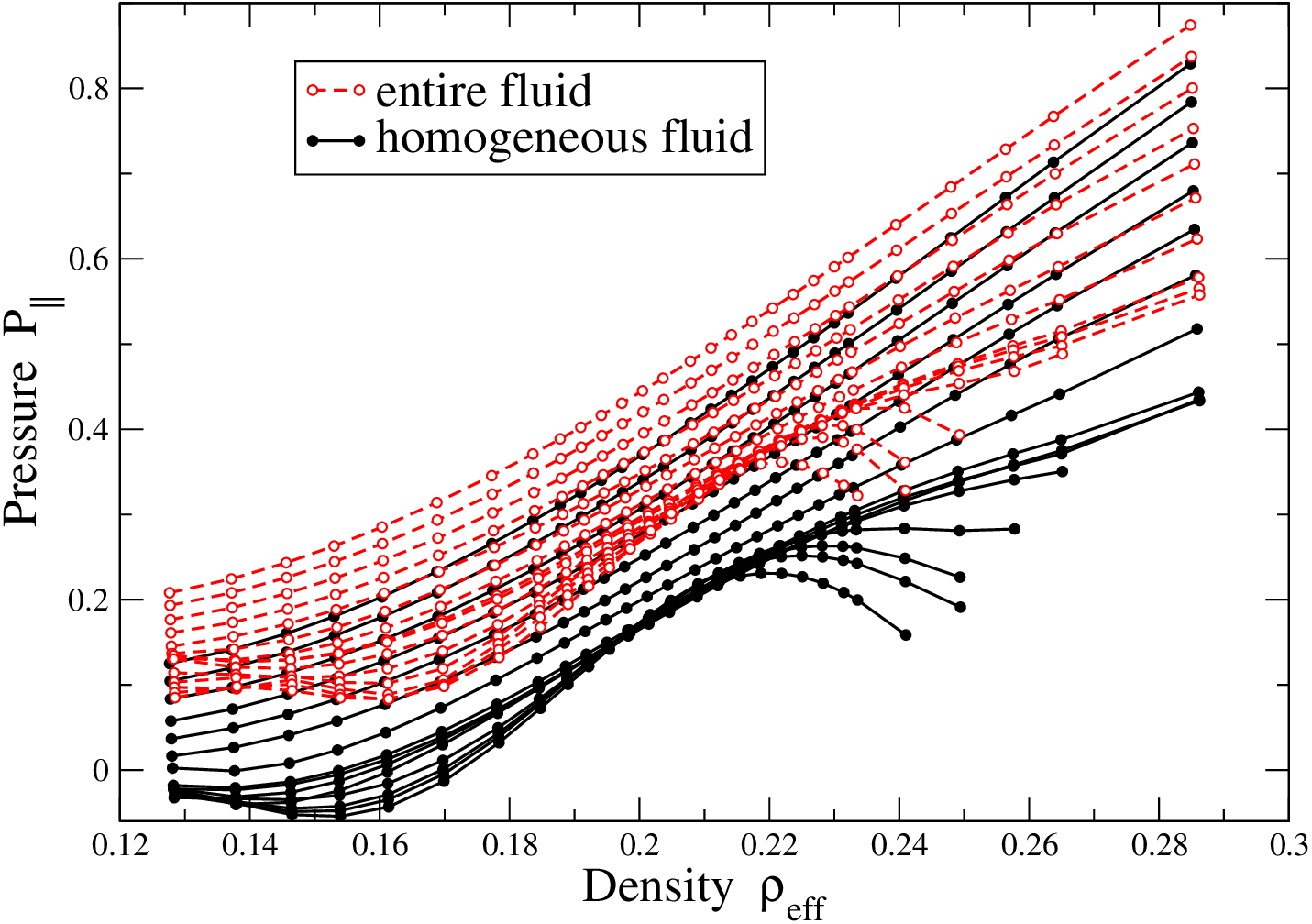}
\vspace{0.3cm}
\caption{\label{fig:2} Comparison of the $P_{\|}-\rho$ phase
  diagram projection for the homogeneous confined fluid 
    (full symbols with continuous
  black lines) and for the entire fluid inside the
  pore (open symbols with dashed red
  lines) including the heterogeneous part. In both sets, isotherms are
from $T=1.4$
  (top) to $T=0.525$ (bottom), as in
Fig.~1b. Isotherms for the entire fluid respect to the homogeneous
fluid are shifted towards higher pressures, especially at low densities.} 
\end{center}
\end{figure}

In the range of temperatures and densities we consider here, fluid
particles in the confined system distribute heterogeneously close to
the walls and homogeneously in the middle of the pore
\cite{leoni2014}. Therefore, a reasonable way to compare the
thermodynamics under confinement with the bulk is to consider only the
contribution coming from the homogeneous confined fluid, as we do in
Section \ref{sec:results}. 

However, in experiments is not easy to separate this contribution from
the signal coming from the overall confined system. Therefore, it
makes sense to ask how the thermodynamic properties would change if we 
include in the calculation the contribution coming from the
heterogeneous part, and how they would compare  with those of the well
defined homogeneous fluid. 

We find that isotherms of the entire fluid are shifted towards
higher pressures with respect to those of the homogeneous fluid,
especially at low densities (Fig.~\ref{fig:2}).  
We understand this effect as a consequence of the density
heterogeneities in the system due to the layering. As seen in Fig.~1 of
Ref.~\cite{leoni2014}, the density profile $\rho(z)$ has large
variations near the walls, especially close to the attractive wall, 
reaching values that are also ten times larger than the average
density. Hence, within this heterogeneous region the inter-particles
distance can be largely reduced with respect to the homogeneous fluid,
inducing large repulsive contributions to the average $P_{\|}$,
defined by the Eq.~(\ref{equ:virial}).

%
%

\section{Conclusions}
\label{sec:conclusions}

We study by MD simulations how the slit pore nano-confinement made of
a structured attractive wall parallel to a smooth repulsive wall
affects the phase diagram and the  anomalies of the CSW fluid and its
water-like properties. 
In a previous work \cite{leoni2014} we found that the slit pore
has a drastic effect on the structural and dynamical properties of this
system. Particles arrange in layers parallel to the walls, especially
near the attractive surface, and for a large range of temperatures and
densities they distribute homogeneously in a central region of the
pore. At low $T$ and high $\rho$ the fluid becomes heterogeneous
(amorphous) and crystallizes near the structured attractive wall.
Interestingly similar results were found in all-atoms simulations of
water under hydrophobic and hydrophilic slit-pore confinement, with
inhomogeneities and ordering near the hydrophilic surface
\cite{Srivastava2011}. 

Here we compare thermodynamic properties of the confined fluid with
the bulk. We consider the homogeneous fluid in the pore and we find
that its isotherms can be rescaled to that of the bulk. This
result implies that the knowledge of the thermodynamics of the
confined fluid can give us a direct insight into the properties of 
the bulk system. In particular, we find that in confinement, as well
as in bulk, the fluid undergoes a liquid-liquid phase transition
ending in a LLCP.  

We study also how the confinement affects the anomalous properties
and, in particular, the TMD line, the diffusion constant and the
structural (translational) order.
We find that the confinement does not change the water-like hierarchy
of anomalies, with the TMD line within the anomalous-diffusion region
and the latter inside the anomalous-structure region, in a cascade of
nested domes.  

However the confinement induces a shift of the external dome towards
larger densities, pressures and temperatures with respect to the
bulk. We interpret this results as a consequence of the enhancement of
structure due to fluid-wall interaction.

We find also a moderate increase of mobility of the homogeneous central
part of the confined fluid. We consider this effect due to the
layering and collective modes of the fluid that are consequence of the
long-range correlations that have been calculated for the CSW
potential. This mechanisms could be relevant also for water under
similar confinement, where layering and hydrogen-bond  long-range
correlation could be the origin of collective modes. 

We observe that our slit-pore confinement promotes the LLCP to
temperatures higher than in bulk. This result could be relevant for
those systems, e.g., water, where the spontaneous crystallization
prevents a direct observation of the LLCP and for which there have 
been recent attempts to measure the LLCP in confinement
\cite{Liu2005, Nagoe:2013ys, Mallamace2014}.
As discussed, we consider that this result is a consequence of the
interaction energy of the fluid with the wall, independent of our
specific choice with competitive surface interactions at the two
parallel walls. In particular, we expect that for a slit pore with
both walls attractive or both repulsive the homogeneous fluid
sub-region would narrow or would broaden, respectively, but for
strong-enough fluid-wall interaction the LLCP temperature would
increase with respect to the bulk case. Further calculations, beyond
the scope of the present work, would be necessary to elucidate this point.

Motivated by understanding if the interpretation of the experiments
with anomalous fluids under confinement could be affected by the
presence of the amorphous (heterogeneous liquid) coexisting with
the homogeneous liquid, we calculate the thermodynamic properties for
the entire pore in absence of the crystal.  
This comparison is useful in those experimental cases in which 
it is difficult to separate the signals associated to the homogeneous
and the heterogeneous liquids.
We find that the isotherms describing the entire fluid with respect to
the homogeneous liquid are shifted towards higher pressures,
especially at low densities. Yet  the physics of the two systems,
homogeneous fluid and homogeneous plus heterogeneous fluid, is
the same, both displaying a GLCP and a LLCP.

Our observations could help in interpreting experiments with confined
anomalous fluids, and especially water, in which it is unclear if the
fluid retains the thermodynamic properties of the bulk
\cite{soper2008,ricci2008}.
However, a limitation to the similarities between confined and bulk
fluid could come from the pore size. Extrapolating the results in 
Ref.~\cite{leoni2014} and the present estimates, we argue that the
LLCP would be observable in a slit pore with a width $>5$~fluid layer,
i.e.  $>1.5$~nm for water, if the walls are 
strongly hydrophobic, or $>10$~layers, i.e. $>3$~nm for water, if the
walls are hydrophilic.  This requirement is satisfied, e.g., by the
hydrophilic Vycor brand porous glass No. 7930 (with average pore size
of 5~nm) used in the experiment of Ref.~\cite{Zanotti2005} to study
the LLCP for hydration water, although in that experiment the
hydration level was only 25\%. 

Nevertheless, our conclusions do not exclude that the LLCP could be 
accessible also in strong hydrophilic confinement, such as in
Ref.s~\cite{Liu2005,Mallamace2007, Zhang2011, Cupane2014,
  CupanePRL2014}, because our study does not consider the
effect of curvature of the confining walls on the crystallization
rate. It is reasonable to hypothesize that the curvature would
diminish the crystallization rate, making the critical region near the
LLCP accessible also in strongly confined water. Further theoretical
study is necessary to clarify this point.

\section*{Acknowledgments}

We thank Marco Bernabei, Valentino Bianco, Carles Calero, David
Reguera and Oriol Vilanova for useful discussions. We acknowledge the
support of Spanish MINECO grant FIS2012-31025.

\bibliography{Leoni_Franzese_resubmission.bib}

\end{document}